\documentstyle[12pt]{article}
\begin{document}
\begin{titlepage}
\ \\
\ \\
\ \\
\ \\
\begin{center}
{\LARGE \bf
Generalized Entropy Composition\\
with Different q Indices:\\
A Trial
}
\end{center}

\begin{center}

 in \\
{\it International Workshop on Classical and Quantum Complexity\\
 and Nonextensive Thermodynamics\\
 (Denton-Texas,  April 3-6, 2000)
}
\end{center}

\begin{center}
\Large{
K.Sasaki and M.Hotta
 }\\
{\it
Department of Physics, Tohoku University,\\
Sendai 980-8578, Japan
}
\end{center}
\ \\
\ \\

\begin{abstract}
We analyze systematically composable composite entropy
of two Tsallis subsystems with different
 q indices. H-theorem and  thermal balance relation
 are commented.
 This report is based mainly on our recent paper \cite{sh}.
\end{abstract}

\end{titlepage}

\section{
Introduction
}
\ \\

 It has already been pointed out that the Tsallis statistical mechanics
 may be useful for explaining many anomalous systems. Nevertheless
 it should be said that its fundamental understanding has not been
 achieved yet, because a lot of crucial problems still remain open.

 For example, why extended entropies take the Tsallis form ?
 What determines its q index ? These topics are certainly a part of
 the most important questions for the formulation.

 On the other hand,
 we believe, it is also an essential problem whether the second law
 of the thermodynamics of composite systems with {\it different} q indices
  holds or not.
  If one of the subsystems takes $q=1$, it
 behaves as an ordinary system and the sub-entropy is
 just Boltzmann-Gibbs. Then increase of the composite entropy
  may be worth guaranteeing that
 the Tsallis form is actually a physically relevant entropy.
 If one takes a small ordinary subsystem compared to the Tsallis
 subsystem and   interaction between the systems is negligibly weak,
  we can regard the ordinary part as a thermometer.
  The setup will enable us to discuss what is an observable temperature
  of the Tsallis subsystem. Significance of the problem was
  first, as far as I know, pointed out exhaustively  by Rajagopal\cite{rj}
  and  a conjecture on
  the thermal balance for the different q-indices case
  is discussed
   by Tsallis\cite{ot}.

 There exist many other  motivations to think composite systems
  with different q indices. For example\cite{plasma},
  non-neutral electron gas
 is often argued as a Tsallis system with $q\sim 0.5$.
 The plasma electrons possess spin degree of freedom
 besides the spatial one. Thus if one wants to
 incorporate the spin thermal fluctuation in the external magnetic field,
 the composite entropy form with both $q_{spatial}\sim 0.5$ and
 $q_{spin}(\neq q_{spatial})$ included will be invoked.
Statistical aspects of
internal degree of freedom (iso-spin, baryon charge, lepton charge and so on)
 of self-gravitating systems may also require such a composite entropy.

In order to write down explicitly the composite entropy form
 with different q indices, guiding principles for the entropy
 should be requested just as symmetries play crucial roles when
 actions of general relativity and quantum field theories are fixed.
 However for the nonextensive statistical mechanics any principles
 are not yet established by physical evidence.
 Thus we must investigate all possibilities that could appear
  in Nature, however, this is too high to access.

Therefore, as a strategy,
we firstly divide all possible composite forms with two categories,
 composable and noncomposable and, as a first step, concentrate on
 the composable entropy category.
 The second category, noncomposable entropy,
 looks technically hard to analyze exhaustively, so we
 keep the non-composable investigation beyond our scope
 of this report.

Tsallis is the first person who
 was aware that the composable property is not automatically equipped and
 emphasized its significance
 in the context of the same-q case, as follows\cite{tsallis}. \\
 {\it It concerns the non-trivial fact
 that the entropy $S(A+B)$ of a system composed
 of two independent subsystems $A$ and $B$ can be calculated from the
 sub-entropies $S(A)$ and $S(B)$, without any need of microscopic knowledge
  about $A$ and $B$, other than the knowledge of some generic universality
   class, herein the nonextensive universality class, represented by
    the entropic index q, i.e., without any knowledge about the microscopic
  possibilities of $A$ and $B$ nor their associated  probabilities.}\\
 Also Joichi and one of the authors\cite{hj}
  demonstrate explicitly powerful ability
 of composability in determination of generalized entropic forms.
For example, uniqueness of the Tsallis entropy has been shown by
imposing composability on a rather generic entropy form as
\begin{eqnarray}
S=C+\sum_i \phi(p_i).
\end{eqnarray}

In this paper, we analyze in detail
composable composite entropy of two Tsallis subsystems
with different q values.
We also show its H-theorem nature and thermal
 balance relation (the zero-th law) in the sense proposed by
 Abe\cite{abe}.
This report is based on a recent work \cite{sh}.
To follow the analysis in more detail, see the original
paper.

\section{Composable Entropy}
\ \\

For the case with the same sub-indices $q$, the Tsallis canonical distribution
 of the composite system is obtained by maximizing the following
 action \cite{tmp}.
\begin{eqnarray}
\tilde{S}&=&S_{A+B}
-\alpha \left( \sum^N_{i=1} \sum^M_{j=1}
P_{ij} -1 \right)
\nonumber\\
&&
-\beta \left(
\sum^N_{i=1} \sum^M_{j=1} E_{ij}P_{ij}
-<E>
\right),\label{1}
\end{eqnarray}
 where $\alpha$ and $\beta$ are Lagrange multipliers and generate
 the unitary condition:
\begin{eqnarray}
\sum^N_{i=1} \sum^M_{j=1} P_{ij} =1,
\end{eqnarray}
 and the total energy constraint:
\begin{eqnarray}
\sum^N_{i=1} \sum^M_{j=1} E_{ij} P_{ij}
=\sum^N_{i=1} \sum^M_{j=1} (E^A_i +E^B_j )P_{ij} =<E>.
\end{eqnarray}
Here $P_{ij}$ denotes the escort function of the probability
$p_{ij}$ as
\begin{eqnarray}
P_{ij} =\frac{(p_{ij})^q }{\sum^N_{k=1} \sum^M_{l=1}
(p_{kl})^q },
\end{eqnarray}
 and the composite entropy $S_{A+B}$ is just given as a standard Tsallis form:
\begin{eqnarray}
S_{A+B}&=&
-\frac{1}{1-q}
\left[
1- \sum^N_{i=1}\sum^M_{j=1} (p_{ij} )^q
\right]
\nonumber\\
&=&
-\frac{1}{1-q}
\left[
1-\left(\sum^N_{i=1} \sum^M_{j=1}
(P_{ij})^{\frac{1}{q} } \right)^{-q}
\right].
\end{eqnarray}
We extend the above formulation in order to include
 the cases with different q sub-indices.
Treating the escort function $P_{ij}$ as the fundamental variable
 for variational procedures admits keeping the Lagrange-multiplier terms
 in eqn (\ref{1}) unchanged,
  because they do not have explicit q dependence. Thus we modify
 the composite entropy form so as to depend on two positive
 sub-indices $q_A $ and $q_B $.

Here let us impose composability on the composite entropy with two indices:
\begin{eqnarray}
S_{A+B} (P_{ij} =P^A_i P^B_j )=\lambda (S_A ,S_B), \label{2}
\end{eqnarray}
where $\lambda$ is arbitrary function of $S_A$ and $S_B$ and
\begin{eqnarray}
&&
S_A =
-\frac{1}{1-q_A }
\left[
1-\left(\sum^N_{i=1} (P^A_i)^{\frac{1}{q_A} } \right)^{-q_A}
\right],
\\
&&
S_B =
-\frac{1}{1-q_B }
\left[
1-\left(\sum^M_{j=1} (P^B_j)^{\frac{1}{q_B} } \right)^{-q_B}
\right].
\end{eqnarray}

Then it can be proven that the most general form satisfying
composability (\ref{2}) is given as
\begin{eqnarray}
S_{A+B} =\Omega (X_a ,\bar{X}_b ;r_A ,r_B) , \label{3}
\end{eqnarray}
where
\begin{eqnarray}
&&
X_1=\sum^M_{j=1}
\left(
\sum^N_{i=1}(P_{ij})^{r_A}
\right)^{\frac{r_B}{r_A}} ,
\\
&&
X_2 =
\sum^M_{j=1}
\left(
\sum^N_{i=1} P_{ij}
\right)^{ r_B } ,
\\
&&
X_3 =\sum^M_{j=1}
\left(
\sum^N_{i=1}(P_{ij})^{r_A}
\right)^{\frac{1}{r_A}} ,
\\
&&
\bar{X}_1=\sum^N_{i=1}
\left(
\sum^M_{j=1}(P_{ij})^{r_B}
\right)^{\frac{r_A}{r_B}} ,
\\
&&
\bar{X}_2 =
\sum^N_{i=1}
\left(
\sum^M_{j=1} P_{ij}
\right)^{ r_A } ,
\\
&&
\bar{X}_3 =\sum^N_{i=1}
\left(
\sum^M_{j=1}(P_{ij})^{r_B}
\right)^{\frac{1}{r_B}} ,
\end{eqnarray}
and
\begin{eqnarray}
&&
r_A =\frac{1}{q_A} ,
\\
&&
r_B =\frac{1}{q_B}.
\end{eqnarray}
Here $\Omega$ is arbitrary function with
permutation symmetry between the sub-systems $A$ and $B$:
\begin{eqnarray}
\Omega (\bar{X}_b ,X_a ;r_B ,r_A)
=
\Omega (X_a ,\bar{X}_b ;r_A ,r_B) .
\end{eqnarray}

Moreover we can introduce another composability when
one constructs a grand
composite system $(A+B)+(A+B)'$ of two composite systems
 $(A+B)$ and $(A+B)'$. Let us impose the following property on their entropies.
\begin{eqnarray}
S_{(A+B)+(A+B)'} =\Lambda (S_{(A+B)}, S_{(A+B)'}) ,\label{4}
\end{eqnarray}
where $\Lambda$ is an arbitrary function. This implies that
the value of
the grand entropy is fixed only by information of the composite entropies
 $S_{(A+B)}$ and $S_{(A+B)'}$.
 Then bi-composability is defined by realization of both the above two
composabilities (\ref{2}) and (\ref{4}). Recall here that
the Tsallis entropy actually satisfies the bi-composability when
the $q$ indices of the sub-systems are the same.
Here we should also stress
that the concept of bi-composability is associated with a set of
the two {\it simultaneous} equations (\ref{2}) {\it and} (\ref{4}),
thus different notion from the original composability, which
 implies a single relation as eqn (\ref{2}), or eqn (\ref{4}).
In fact some composable entropy forms which satisfy
 eqn (\ref{2}) or eqn (\ref{4}) do {\it not} show
 the bi-composability
 even when the subsystems are statistically independent.
Also note that the there is no reason in general that
the functional form $\Lambda$ coincides with the form of $\lambda$.

It is  possible to write down the most generic form of the bi-composable
 entropy and the result is as follows.
\begin{eqnarray}
S_{A+B} =F(\Delta ; r_A ,r_B), \label{5}
\end{eqnarray}
where $F(x ;r_A ,r_B)$ is an arbitrary function and
\begin{eqnarray}
\Delta =\prod^3_{a=1} (X_a)^{-\nu_a}
\prod^3_{b=1} (\bar{X}_b)^{-\bar{\nu}_b} >0.
\label{5}
\end{eqnarray}
Here the exponents $\nu_a$ and $\bar{\nu}_b$ are arbitrary constants
in this level and
 supposed to be fixed by dynamical property of the system.

We can propose a simple and attractive toy model for the
 bi-composable entropy. If one assumes the Tsallis-type nonextensivity:
\begin{eqnarray}
S_{(A+B)+(A+B)'}=S_{(A+B)} +S_{(A+B)'} +(1-Q) S_{(A+B)}S_{(A+B)'} \label{6}
\end{eqnarray}
for the grand composite system,
it is shown that the entropy must take the following form.
\begin{eqnarray}
S_{A+B} =-\frac{1-\Delta}{1-Q} .\label{7}
\end{eqnarray}
Here $Q$ behaves as a grand index of the composite system and is expected
 to be determined by some dynamical information of the system,
 just like usual q index. Later we call the simple model (\ref{7})
 Tsallis-type bi-composable entropy.

\section{H-Theorem}
\ \\

Here we comment on H-theorem for the composable entropy.
Unfortunately H-theorem does not hold in a strong sense
 for all the composable entropies in eqn (\ref{3}).
 It is proven analytically that there exists a master equational
 dynamics in which some probability configurations give negative
  values of time-derivative of the composite entropies.
 However this fact may not be so significant
  for real physical systems,
 because the fixed-two-subindices picture does not always need to work
  when the total system is composed of
  two Tsallis systems with {\it different} q indices.
  Each original subindex is widely
  believed to be chosen dynamically for each isolated system.
  Thus it may happen in general that
  the interaction between the two subsystems drives the q values changed,
   or the q-deformed statistical picture itself gets broken
   and should be replaced by more microscopical pictures.
  Therefore it sounds plausible that the physical situations
  are somehow limited in which the total system can be regarded as
 a composite system of two independent q-deformed subsystems.
Meanwhile it seems natural, at least, to consider that
  the picture should work well
  when the interaction between the subsystems are negligibly weak,
  or the two subsystems are in near-equilibrium states.
Actually H-theorem for such physically relevant cases can be  exactly
proven for a part of the bi-composable entropies.
The Tsallis-type bi-composable entropies (\ref{7}) which satisfy
\begin{eqnarray}
&&
r_B (r_B -r_A ) \nu_1 +r_B (r_B -1) \nu_2 -(r_A -1)\nu_3 =0 ,\label{8}
\\
&&
r_A (r_A -r_B ) \bar{\nu}_1
+r_A (r_A -1) \bar{\nu}_2 -(r_B -1)\bar{\nu}_3 =0 \label{9}
\end{eqnarray}
 and
\begin{eqnarray}
\frac{
r_B(r_A -1) \nu_1 +r_A (r_B -1) \bar{\nu}_1
+(r_A -1)\nu_3
+(r_B -1)\bar{\nu}_3
}{1-Q} >0 \label{10}
\end{eqnarray}
 do not decrease in time for the near-microcanonical-equilibrium case:
\begin{eqnarray}
P_{ij} (t) = \frac{1}{NM} +\epsilon_{ij} (t),\label{12}
\end{eqnarray}
where $\epsilon_{ij}$ is infinitesimal deviation from the
 equipartition distribution.
 It is also noticed that for the negligibly-weak interaction case:
\begin{eqnarray}
P_{ij} (t) = P^A_i (t) P^B_j (t), \label{11}
\end{eqnarray}
 the Tsallis-type bi-composable entropy satisfying (\ref{8}),(\ref{9}) and
 (\ref{10}) preserves H-theorem. It is proven that if
\begin{eqnarray}
&&
c_A =\frac{1-q_A}{1-Q}
\left(
r_B \nu_1 +r_A \bar{\nu}_1
+r_A \bar{\nu}_2 +\nu_3
\right) >0 ,\label{002}
\\
&&
c_B =\frac{1-q_B}{1-Q}
\left(
r_B \nu_1 +r_A \bar{\nu}_1
+r_B \nu_2 +\bar{\nu}_3
\right)>0 \label{003}
\end{eqnarray}
 hold, the entropy has non-negative time-derivative for the evolution
 (\ref{11}). Using the relations (\ref{8}) and (\ref{9}), the conditions
 (\ref{002}) and (\ref{003}) can be rewritten into
 the same condition and that is just the third relation (\ref{10}).

It is worth noting that for a rather general bi-composable entropy
defined by use of an arbitrary function $G(x)$ monotonically increasing  as
\begin{eqnarray}
S_{A+B} =G\left( -\frac{1-\Delta}{1-Q} \right),
\label{130}
\end{eqnarray}
the H-theorem still holds in the cases (\ref{12}) and (\ref{11})
 if the relations (\ref{8}), (\ref{9}) and (\ref{10}) are
 simultaneously satisfied.
 For the entropy (\ref{130}),  we discuss next
 Abe's thermal balance relation \cite{abe}, that is,
 the zero-th law of the thermodynamics.

\section{Thermal Balance Relation}
\ \\

It turns out that
the nonextensivity of the bi-composable entropy (\ref{130})
is given as follows.
\begin{eqnarray}
S_{A+B} &=& G\left( -\frac{1-\bar{\Delta}}{1-Q} \right), \nonumber\\
\bar{\Delta} &=&
\left[
1+(1-q_A ) S_A
\right]^{r_B \nu_1 +\nu_3 +r_A \bar{\nu}_1 +r_A \bar{\nu}_2 }
\nonumber\\
&&\times
\left[
1+(1-q_B ) S_B
\right]^{r_A \bar{\nu}_1 +\bar{\nu}_3 +r_B \nu_1 +r_B \nu_2 } .
\end{eqnarray}
Using the H-theorem conditions (\ref{8}) and (\ref{9}),
the variation of the entropy can be written as
\begin{eqnarray}
\delta S_{A+B}
&=&\frac{r_B (r_B -1) (\nu_1 +\nu_2 )
+r_A (r_A -1) (\bar{\nu}_1 +\bar{\nu}_2 ) }{1-Q}\Delta G'
\nonumber\\
&&
\times
\left[
\frac{q_A}{1+(1-q_A) S_A}\delta S_A
+
\frac{q_B}{1+(1-q_B) S_B}\delta S_B
\right] .
\end{eqnarray}
Along Abe's argument for the same q case\cite{abe}, we take $\delta S_{A+B}=0$
under the total energy conservation relation:
\begin{eqnarray}
\delta E_A +\delta E_B =0
\end{eqnarray}
to get the thermal balance relation. The procedure is expected valid
 for thermodynamic-limit situations as in the same q case.
The result is given as follows,
 independent from the functional form $G(x)$ and the value of $Q$.
\begin{eqnarray}
\frac{q_A}{1+(1-q_A) S_A}\frac{\delta S_A}{\delta E_A}
=
\frac{q_B}{1+(1-q_B) S_B}\frac{\delta S_B}{\delta E_B} .\label{13}
\end{eqnarray}
Note that this relation includes the Abe's balance relation \cite{abe}
 as a special case.
Actually when $q_A =q_B =q$ is taken
\begin{eqnarray}
\frac{1}{1+(1-q) S_A}\frac{\delta S_A}{\delta E_A}
=
\frac{1}{1+(1-q) S_B}\frac{\delta S_B}{\delta E_B}
\end{eqnarray}
is exactly reproduced. Also eqn (\ref{13}) is
 consistent with a guessed relation \cite{ot}
 by Tsallis for the different q case.

If the system $A$ is taken as an ordinary Boltzmann-Gibbs system
($q_A =1$), the sub-entropy is reduced into the BG form $S_{BG:A}$.
Then the relation
(\ref{13}) with $q_A =1$ is expressed as
\begin{eqnarray}
\frac{\delta S_{BG:A}}{\delta E_A}
=
\frac{q_B}{1+(1-q_B) S_B}\frac{\delta S_B}{\delta E_B}.
\end{eqnarray}
Here it is a trivial fact that physical temperature $T_{phys}$ can be
introduced for the system $A$ as follows.
\begin{eqnarray}
\frac{1}{T_{phys}}=\frac{\delta S_{BG:A}}{\delta E_A}.
\end{eqnarray}
Therefore observable temperature $T_B$ of the Tsallis system $B$ should be
defined as
\begin{eqnarray}
\frac{1}{T_B}=\frac{q_B}{1+(1-q_B) S_B}\frac{\delta S_B}{\delta E_B},
\label{150}
\end{eqnarray}
so as to preserve the zero-th law of thermodynamics:
\begin{eqnarray}
\frac{1}{T_{phys}} =\frac{1}{T_B}.
\end{eqnarray}
Here we should stress that before our analysis
no one argues explicitly presence of the numerator $q_B$ of the prefactor
 in the right-hand-side term of eqn (\ref{150}).
Due to the definition (\ref{150}), the
original relation (\ref{13}),  in which $q_A$ is not needed to take unit,
can be interpreted as a generalized thermal balance as follows.
\begin{eqnarray}
\frac{1}{T_A} =\frac{1}{T_B}.\label{14}
\end{eqnarray}
The transitivity relation (\ref{14}) looks quite plausible
 and attractive, though the derivation remains still heuristic.

\section{Final Remarks}
\ \\

We discussed Tsallis entropy composition with different q indices
 in detail and showed several rigid conclusions for
 the composable entropy class.
 However we must agree that there still remain open problems.
 For example, what determines the functional form $\Omega$,
 or $F$, and  the exponents $\nu_a $, $\bar{\nu}_b$ for the bi-composable
 entropy and the grand index $Q$ for the Tsallis-type model ?
 How calculate them ? These questions
 are perhaps as profound and difficult as the original q index problem
 of the Tsallis entropy.
We just expect that they are determined by dynamical and
 somewhat microscopical information of  physical systems.

However it is possible, as seen below, just to
construct a simple and regular model as a special solution of the H-theorem
 problem.
\begin{eqnarray}
S_{A+B}
&=&
-\frac{r_A (r_A -1)^2 +r_B (r_B -1)^2}
{(r_A -1)(r_B-1) (r_A +r_B -2) }
\nonumber\\
&&\times
\left[
1-
\left(
X_1^{r_B -1} \bar{X}_1^{r_A -1}
\left[
\frac{X_2}{\bar{X}_2}
\right]^{r_A -r_B}
\right)^{
\frac{2-r_A -r_B }
{2\left[
r_A (r_A -1)^2 +r_B (r_B -1)^2
\right]}
}
\right].\label{15}
\end{eqnarray}
This example succeeds in simplification of the form
 because it does not depend on $X_3$ and $\bar{X}_3$, while the H-theorem
 still holds.
Also it has regular limits for both $r_A \rightarrow 1$ and  $r_B\rightarrow 1$
 independently, and no singularities for positive $q_A$  and $q_B$ region.
(The limit $r_A +r_B \rightarrow 2$ is also regular.)
Moreover it is  easily confirmed that
if one takes $q_A =q_B $, the form is reduced into the original Tsallis form.
This toy model may be useful for future works to get more deeper
intuition about physics of the generalized entropy composition.

Finally we want to comment that
our analysis has not included at all the class of
 noncomposable composite entropy and
it may be an interesting open problem to analyze the case.

\end{document}